\documentclass[pra,preprint,showpacs,a4paper,floatfix,nofootinbib]{revtex4}

\usepackage[dvips]{graphicx}

\usepackage{amsmath}
\usepackage{amsfonts}
\usepackage{amssymb}

\begin{document}

%%%%%%%%%%%%%%%%%%%%%%%%%%%%%%%%%%%%
% shorthand notations which allow to easily switch between
% Kostyuk's notations and mine:
%
%  command            my notation    Kostyuk's notation
\newcommand {\ls}    {{s}}       %   {{l}}
\newcommand {\ps}    {{p_s}}       %   {{p_l}}
\newcommand {\dotl}  {{\dot{s}}}   %   {{\dot{l}}}
\newcommand {\ddotl} {{\ddot{s}}}  %   {{\ddot{l}}}

\newcommand {\ku}  {{k_{\rm u}}}
\newcommand {\lambdau}  {{\lambda_{\rm u}}}
\newcommand {\gammar}  {{\gamma_{\rm r}}}

%%% My standard commands
\newcommand {\Om}  {{\Omega}}
\newcommand {\om}  {{\omega}}
\newcommand {\E}  {{\varepsilon}}
\newcommand {\dUmax} {U^{\prime}_{\max}}

\newcommand {\Ld} {{L_{\rm d}}}
\newcommand {\La} {{L_{\rm a}}}

\newcommand {\calE}  {{\cal E}}
\newcommand {\calJ}  {{\cal J}}
\newcommand {\calEb}  {{\cal E}_{\rm b}}

\newcommand {\atf}  {{a}_{\rm TF}}
\newcommand {\opt}  {{\rm opt}}

\newcommand {\bfE}  {{\bf E}}
\newcommand {\bfn}  {{\bf n}}

\newcommand {{\mg}} {{\mathfrak{g}}}

\newcommand{\sign}{\mbox{sign}}

\newcommand{\ds}{\displaystyle}

%%%%%%%%%%%%%%%%%%%%%%%%%%%%%%%%%%%%%%%%%%%%%%%%
\title{One-dimensional Model of a  Gamma Klystron}

\author{Andriy Kostyuk$^{1,2}$, Andrei Korol$^{1,3}$, Andrey Solov'yov$^{1}$, and Walter Greiner$^{1}$}

\affiliation{ %\phantom{a}\\
$^1$ Frankfurt Institute for Advanced Studies, Johann Wolfgang G\"othe-Universit\"at,
Ruth-Moufang-Str. 1, 60438 Frankfurt am Main, Germany \\
$^2$ Bogolyubov Institute for Theoretical Physics,
      Metrologichna str. 14b, 03143 Kyiv, Ukraine \\
$^3$ Department of Physics, St. Petersburg State Maritime 
Technical University, Leninskii prospect 101,
St. Petersburg 198262, Russia}

%\date{\today}

\begin{abstract}
A new scheme for amplification of coherent gamma rays is proposed.
The key elements are crystalline undulators ---
single crystals with
periodically bent crystallographic planes exposed to a high energy 
beam of charged particles undergoing channeling inside the crystals.
The scheme consists of two such crystals separated by a vacuum gap. The beam 
passes the crystals successively. The particles
perform undulator motion inside the crystals following the periodic shape of
the crystallographic planes.
Gamma rays passing the crystals parallel to the beam get amplified
due to interaction with the particles inside the crystals. The term
`gamma klystron' is proposed for the scheme because its operational principles
are similar to those of the optical klystron. A
more simple one-crystal scheme is considered as well for the sake of comparison. It is
shown that the gamma ray amplification in the klystron scheme can be
reached at considerably lower particle densities than in the
one-crystal scheme, provided that the gap between the crystals is
sufficiently large.
\end{abstract}

\pacs{41.60.-m, 61.85.+p}

\maketitle
%%%%%%%%%%%%%%%%%%%%%%%%%%%%%%%%%%%%%%%%%%%%%%%%%%%%%%%%%%%%%%%%%%%
\section{Introduction}

Development of coherent radiation sources for a wavelength below
0.1 nm (which corresponds to the photon energy of tens keV or
higher), i.e. in the gamma ray range, is a challenging goal of
modern physics.
%Sub-angstrom wavelength coherent
Such radiation will
have many applications in
 basic science, technology and medicine.
In particular, they will have a revolutionary impact in nuclear and
solid state physics as well as in life sciences.

The present state-of-the-art lasers are capable for emitting
electromagnetic radiation from the infrared to ultraviolet range of
spectrum. X-ray free-electron lasers 
are currently under construction \cite{SLAC,DESY}. Moving further,
i.e. into gamma ray band, is not possible without new approaches and
technologies.

%\cite{Ginzburg1947,Madey1971,Martellucci1983,Marshall1985,LaserHandbook1985,LuchiniMotz1990,Brau1990,LecturesDattoli1993,Freund1996,Ciocci2000,Saldin2000}

One of the most promising approaches utilizes the spontaneous emission 
of gamma rays by ultrarelativistic charged particles undergoing channeling 
in periodically bent crystals. An undulator-type radiation is emitted in such a system
in addition to the well-known channeling radiation due 
to the periodic motion of the particle that follows the shape of crystallographic 
planes or axes.
The feasibility of gamma ray generation by ultrarelativistic positrons 
in planar channeling regime  in crystals with periodically bent 
crystallographic planes
was proven a decade ago \cite{KSG1998,KSG1999}
(see also the review \cite{KSG2004_review} and  references therein).
Such a device is known as a 'crystalline undulator'.\footnote{The term
'crystalline undulator' was introduced in
\cite{KaplinPlotnikovVorobev1980} but was not elaborated there.}
Recently, the feasibility of the crystalline undulator utilizing the
planar channeling of electrons 
was demonstrated \cite{PRL,TKSG21007_jpg}.

The operation of the crystalline undulator is based on the phenomenon 
of charged particle channeling in crystals \cite{Lindhard1965}
(see also the latest review \cite{Uggerhoj2005}).
Channeling takes place if a charged particle enters a crystal
nearly parallel to a crystallographic axis or plane. It can be
confined by the electrostatic potential of crystal atoms so that it
moves along the axis or plane.  It is remarkable that the particle
follows the shape of the axis or the plane if they are bent. Therefore 
the particle trajectory inside the crystal can be controlled by
choosing a suitable shape of the axes or planes \cite{Tsyganov1976}.

If crystallographic axes and planes  are
periodically bent,
the trajectories of channeling particles are
approximately sinusoidal.  This results into emission of
electromagnetic radiation predominantly in forward direction
\cite{KSG1998,KSG1999,KSG2004_review}.
The process of emission is very similar to that in the periodic 
magnetic field
of an undulator \cite{Ginzburg1947,Motz1951,Motz1953}. 
However,
the electrostatic fields inside a crystal are so strong that they
are able to steer the particles much more effectively than even the
most advanced superconductive magnets.\footnote{Indeed,  the
interplanar electrostatic field in crystals is typically of the
order of 10 GeV/cm which is equivalent to the magnetic field of
approximately 3000 T. The present state-of-the-art superconductive
magnets produce the magnetic flux density of the order of 10 T
with 45 T being currently the highest value obtained by
combining  superconductive and resistive magnets \cite{45Tmagnet}. }
Due to this fact, the
period of crystal bending can be made two to four orders of
magnitude smaller than the period of a conventional 
undulator.\footnote{Periodical bending can be obtained
by the propagation of an acoustic wave along the crystal 
\cite{KSG1998,KSG1999} or  by 
making regularly spaced grooves on the crystal surface 
either by a diamond blade \cite{BellucciEtal2003,GuidiEtAl_2005}
or by means of laser-ablation \cite{Uggerhoj2006_Connell2006}.
Another method is a deposition
of Si$_3$N$_4$ layers onto the surface of a Si crystal \cite{GuidiEtAl_2005}.
One more possibility is growing of Si$_{1-x}$Ge$_x$ crystals\cite{Breese97}
with a periodically varying Ge content $x$ \cite{MikkelsenUggerhoj2000,Darmstadt01}.
} 
Therefore the particle in a periodically bent crystal will emit
electromagnetic radiation with much shorter wavelength, i.e. in the
gamma ray frequency range.

The emission of the gamma rays will become even more
powerful if the positions of
the charged particles in a crystalline undulator are correlated
along the direction of their motion in such a way that the waves
emitted by all particles have approximately the same phase.  The
resulting radiation is expected to be not only more powerful but
also more coherent than the spontaneous radiation from uncorrelated particles.
The physical picture of the coherent emission by
correlated beam is similar to what happens in free electron
lasers \cite{Madey1971,LuchiniMotz1990,RullhusenArtruDhez1998,Saldin2000}. 
Therefore, we  use for this process the term  'lasing effect
in a crystalline undulator'.

First, we consider a one-crystal scheme. The feasibility of 
obtaining a lasing effect in one crystalline undulator  has
been studied in \cite{KSG1999} within the quantum formalism.  
In the current work, this problem is revisited within the classical approach.
Indeed,
the emission of comparatively low energy photons by a  high energy
%($ \E>10^2$ MeV)
particle  (i.e., when  $\hbar \om/ \E \ll 1$, with $\hbar \om$ and
$\E$ being, respectively, the photon and the particle energy) can be
treated classically. The classical approach \cite{Colson1977} has
been widely used for the description of conventional free electron
lasers (see, for instance \cite{LuchiniMotz1990,Saldin2000} and
references therein).

Numerical estimation of the 
parameters of the positron-based crystalline undulator confirm the
results of \cite{KSG1999,KSG2004_review}: extremely high
densities of channeling positrons ($\sim 10^{21}$ cm$^{-3}$ or higher)
would be needed to obtain a lasing effect in one periodically
bent crystal. This suggest the necessity to search for alternative 
schemes which would be capable of reaching the lasing effect at considerably 
lower particle densities. 
 
Reconsideration of the one-crystal scheme was necessary to develop the classical
formalism applicable for the description of the lasing effect in 
crystalline undulator in presence of the
dechanneling phenomenon (i.e. the process of leaving the channeling regime due
to scattering by crystal electrons and nuclei).  
This formalism is more convenient than the quantum one for the developing of 
alternative schemes. Additionally, the one-crystal scheme, being the simplest one,
provide a good reference point for the comparison with more sophisticated set-ups. 

We propose a two-crystal scheme in which the periodically bent
crystals separated by a vacuum gap. The beam of charged particles
passes both crystals successively. In the first crystal,
a correlation is achieved  between the momentum of the particle and its
position along the beam direction. This correlation is  transformed to a density
modulation of the beam. The gamma radiation is generated mostly in
the second crystal by the density-modulated beam. 
This scheme is similar to that of the optical klystron
\cite{Vinokurov,VinokurovSkrinsky}, but is expected to operate in
the gamma ray range.  Therefore we call it a `gamma klystron'.

The article is organized as follows. In section \ref{one-crystal-amp},
a one dimensional model of the one-crystal gamma ray amplifier is considered,
the formalism is introduced, and numerical estimation are presented. 
The results of this section are used as a basis for the comparison in section 
\ref{two-crystal-amp}, where the main subject of our work --- the gamma klystron ---
is considered. The results are summarized and discussed in section \ref{conclussion}.

%This includes the classical 

%%%%%%%%%%%%%%%%%%%%%%%%%%%%%%%%%%%%%%%%%%%%%%%%
\section{One-crystal gamma ray amplifier} \label{one-crystal-amp}

%%%%%%%%%%%%%%%%%%%%%%%%%%%%
\begin{figure}[bht]
\includegraphics[width=\textwidth]{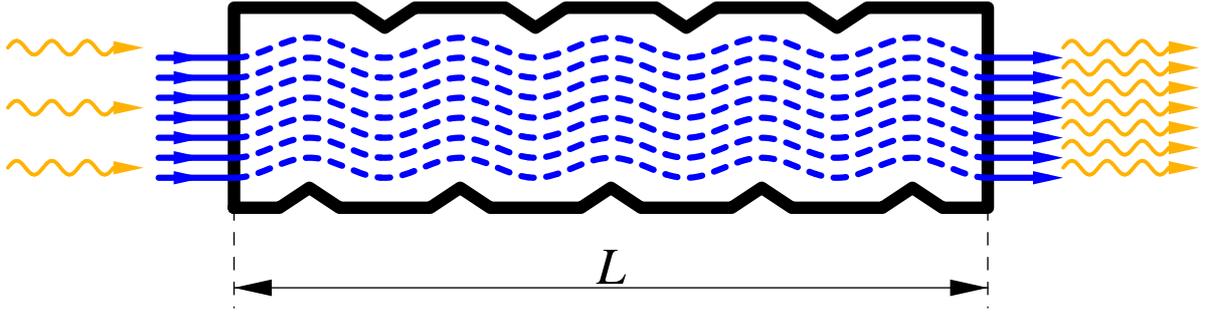}
\caption{A scheme of the one-crystal amplifier. A charged particle
beam (solid lines) and initial (seed) radiation (solid wavy lines)
enter a crystal with periodically bent crystallographic planes.
The particle follow the shape of the crystallographic planes and move
along nearly sinusoidal trajectories (wavy dashed lines).
The radiation is amplified due to its interaction with the beam
in the crystal (see also explanations in the text). }
\label{amplifier}
\end{figure}
%%%%%%%%%%%%%%%%%%%%%%%%%%%%

As it has been already mentioned above,
the lasing effect in a crystalline undulator takes place if
the positions of the
channeling particles along the beam direction are correlated
in such a way that the electromagnetic waves emitted by all particle have
approximately the same phase. This is accomplished by a spatial
modulation (termed usually as 'bunching'
\cite{LuchiniMotz1990,RullhusenArtruDhez1998} or 'microbunching'
\cite{SLAC}) of the particle density along the direction of the beam
motion with the period equal to the wavelength of the emitted
radiation. To obtain such a modulation, initial (seed) radiation
from an external source is needed. This radiation may be generated 
by spontaneous emission of charged particles in a crystalline 
undulator or in the field of a infrared laser wave. In both cases 
the initial radiation has to be well collimated to ensure sufficient 
monochromaticity and coherence.

Under certain conditions which are discussed below, this radiation
modulates the density of the particles channeling in a periodically
bent crystal. Then the bunched beam produces additional radiation of the
same wavelength.

In other words, the undulator amplifies the initial radiation. The
amplifier is the only possible type of lasers in the hard x-ray and
gamma ray range.\footnote{ In the optical range,  the output
radiation feedback is commonly used in free electron lasers instead
of an external source. The feedback is carried out by the mirrors of
an optical resonator. Such class of devices are called 'oscillators'
\cite{Saldin2000}. This approach cannot be applied  in the hard x-ray
or gamma-ray range because of the absence of the appropriate mirrors.}

The scheme of the gamma ray amplifier based on one crystalline
undulator (we call it `one-crystal gamma ray amplifier') is shown in
figure \ref{amplifier}. A charged particles beam and initial gamma
radiation fall parallel to each other  onto a periodically bent
crystal of the length $L$. The beam is aligned with
the tangent of the bent crystallographic planes at the entrance . Therefore, the
particles are captured in the channeling mode and move along the
planes following their shape. In this section we study the
conditions under which the interaction between the radiation and the
channeling particles leads to amplification of the radiation.

%%%%%%%%%%%%%%%%%%%%%%%%%%%%%%
\subsection{Particle dynamics \label{dynamics}}

Let us consider a one-dimensional model of the gamma ray
amplifier.
We assume the crystal and the beam to be infinitely wide in
the $x$ and $y$ directions, which are perpendicular to the direction of the
 beam propagation (the $z$ axis).

\begin{figure}[tb]
\includegraphics[width=\textwidth]{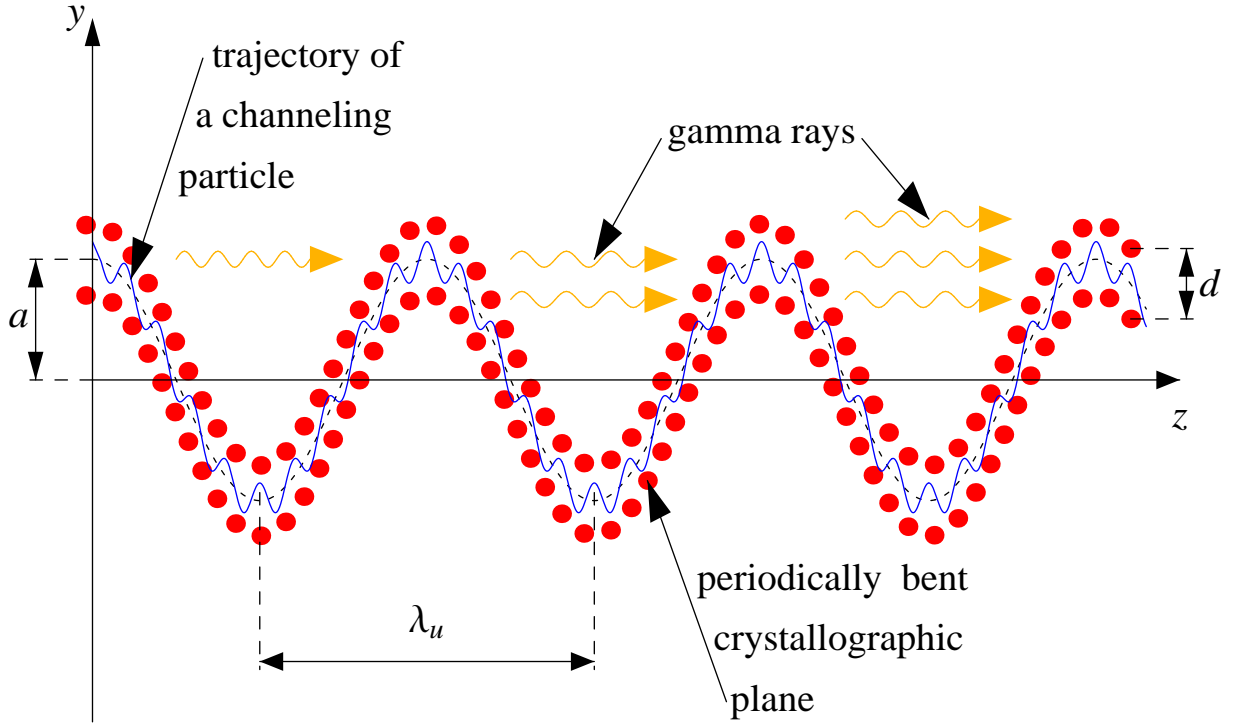}
\caption{A positively charged particle trajectory inside a
periodically bent crystal. Due to the repulsive potential of the
atomic nuclei (filled circles) the particle gets confined inside the
interplanar channel of the width $d$ and moves along the bent
crystallographic planes following their nearly sinusoidal shape with
the period $\lambda_{\rm u}$ and the amplitude $a$ and emitting
electromagnetic waves. Note that the scheme does not reflect the relative 
scale of $d$, $a$ and $\lambda_{\rm u}$ adequately. 
In reality, these quantities satisfy the following double inequality 
$d \ll a \ll \lambda_{\rm u}$.}
\label{undulator}
\end{figure}

The crystallographic planes are parallel to the $x$ axes and are
bent in $y$-direction, so that particles channel in $z$-direction
and oscillate along the $y$-axis (see figure \ref{undulator}).
%%%%%%%%%%%%%%%%%%%%%%%%%%%%%%%%%%%%%%%%%%%%%%%%
%%%%%%%%%%%%%%%%%%%%%%%%%%%%%%%%%%%%%%%%%%%%%%%%

In what follows the shape  $y(z)$ of the periodically bent midplane
is assumed to have a harmonic form:
\begin{equation}
y = a \cos (\ku z),
\label{dynamics.1}
\end{equation}
where $\ku = 2\pi/\lambdau$ with $\lambdau$ being the undulator period.

The magnitude of the undulator period exceeds greatly the bending
amplitude,  i.e. $\lambdau \gg a$ (or, equivalently, $\ku a \ll 1$)
\cite{KSG2004_review}. This inequality, which implies that the
crystal deformation is an elastic one, follows from the important
condition for a stable channeling in a periodically bent crystal.
Stable channeling occurs if the maximum centrifugal force $F_{\rm
cf}$ in the channel is less than the maximum interplanar force
$\dUmax$, i.e. $C\equiv F_{\rm cf}/\dUmax < 1$. For an
ultra-relativistic particle, the centrifugal force depends on the
particle energy as
 $F_{\rm cf}\approx \E/R_{\min}$, where
$R_{\min}=\lambda_{\rm u}^{2}/4\pi^{2}a$ is the minimum curvature
radius of the channel with the bending profile given by
(\ref{dynamics.1}). Hence, the condition reads
\cite{KSG1998,KSG1999}
\begin{eqnarray}
C = \frac{4\pi^2\E a}{\dUmax \lambda_{\rm u}^2} < 1.
\label{dynamics.2}
\end{eqnarray}
A thorough analysis carried out for a number of crystals
\cite{KSG2004_review} allowed one to conclude that the crystalline
undulator is feasible for the values of the bending parameter $C$
lying within the interval $0.01\dots 0.2$. For these $C$ values one
obtains $a \sim 10^{-5} \dots 10^{-3}\, \lambdau$.

Another important feature of a crystalline undulator concerns the
relative magnitude of the bending amplitude $a$ and the interplanar
spacing $d$. It was demonstrated (see
\cite{KSG1998,KSG1999,KSG2004_review}) that the operation of a
crystalline undulator should be considered in the large-amplitude
regime, which implies $d  \ll a$ (typically, $a \sim 10\dots10^2
d$). This condition, in particular, allows one to neglect the
oscillations of the particle in the channel. Thus, one can assume
that under the action of the interplanar field,  the particle
follows the trajectory (\ref{dynamics.1}). Formally this means that
the length $s$, measured along the trajectory, can be used as a
generalized coordinate which uniquely characterizes the position of
the particle. The corresponding conjugate variable is the tangent
projection of the particle momentum,
\begin{eqnarray}
\ps = \frac{m \dotl}{\Bigl(1 - \dotl^2/c^2\Bigr)^{1/2}},
\label{dynamics.3}
\end{eqnarray}
where $m$ is the mass of the projectile and $\dotl\equiv d \ls/d t$.

The evolution of $\ps$ is due to the presence of the amplified
radiation whose intensity is $\bfE$. The equation of motion reads
 \begin{equation}
\frac{d \ps}{d t} = e (\bfE \bfn),
\label{dynamics.4}
\end{equation}
where $e$ is the charge of the projectile, and $\bfn$ is the unit
tangent vector, $\bfn = \Bigl( 0, y_z^{\prime}, 1
\Bigr)/\sqrt{1+y_z^{\prime\,2}}$ (where $y_z^{\prime} = d y/d z
\propto \ku a \ll 1$).

The amplified radiation is sought in the form of a plane wave linearly
polarized along the
$y$ direction and characterized by  the wavelength $\lambda$ and the frequency
$\om = c k$ (here $k = {2 \pi/\lambda}$ and $c$ is the speed of light).
Neglecting the attenuation\footnote{It can be shown that the attenuation length $L_a$ of gamma rays
and hard X rays (see figure 6 in \cite{KSG2004_review}) is by approximately an order of magnitude larger
than the dechanneling length $L_d$ that determines the optimum length of the crystalline undulator.} of the 
radiation in the crystal, one describes
the only non-zero component of the electric field of the amplified wave as follows
\begin{equation}
E_y = E_0 \cos(k z - \om t + \phi) \,.
\label{dynamics.5}
\end{equation}

%%%%%%%%%%%%%%%%%%%%%%%%%% Figure acceldecel.eps
%% !!! This figure was not cited in the Kostyuk's text!!!
%\begin{figure}[tb]
%\includegraphics[width=\textwidth]{acceldecel.eps}
%\caption{
%{\bf Korol: to edit the figure + caption.}
%{\it Particle acceleration and deceleration along a sinusoidal
%trajectory in presence of an electromagnetic wave.}}
%\end{figure}
%%%%%%%%%%%%%%%%%%%%%%%%%%%%%%%%%%%

Substituting (\ref{dynamics.5}) into (\ref{dynamics.4})
 and taking into account that $a k_{\rm u} \ll 1$ one derives
%\begin{equation}
%{d \ps \over d t} = - {e E_0 a \ku \over 2}
%\left[
%\sin \Bigl( (k+\ku) z - \om t + \phi \Bigr)
%-
%\sin \Bigl( (k-\ku) z - \om t + \phi \Bigr)
%\right].
%\label{dynamics.6}
%\end{equation}
%The coordinate $z$ is related to the length of the sinusoidal
%particle path $l$ via
%\begin{equation}
%z = \varkappa l + \frac{1}{8} a^2 k_{\rm u} \sin (2 k_{\rm u} \varkappa l)
%+ a O \left[ (a k_{\rm u})^3 \right],
%\end{equation}
%Then (\ref{dynamics.6}) can be rewritten as
\begin{eqnarray}
\frac{d \ps}{d t}
&=& - \frac{e E_0 a \ku}{2}
\left[
\sin \left( \psi + \frac{a^2 \ku (k+\ku)}{8} \sin (2 \ku \varkappa \ls) \right)
\right.
\nonumber\\
& &
\qquad\qquad
\left.
-
\sin\left( \psi + \frac{a^2 \ku (k-\ku)}{8}
\sin (2 \ku \varkappa s) -2 \ku \varkappa \ls \right)
\right],
\label{dynamics.7}
\end{eqnarray}
where the following notations are used
\begin{eqnarray}
&&\psi = (k+k_{\rm u}) \varkappa \ls - \om t + \phi,
\label{dynamics.8}\\
&&\varkappa =  1 - \frac{(a k_{\rm u})^2}{4}.
\label{dynamics.9}
\end{eqnarray}

The energy exchange between the particle and the electromagnetic
field is most effective when the phase $\psi$ stays nearly
constant, otherwise the first sine term on the right-hand
side of (\ref{dynamics.7}) oscillates, thus averaging out the energy
exchange \footnote{The phase of the second sine term cannot be made
constant. Therefore, this term always oscillates. Hence the main
contribution to the energy exchange is due to the first sine term.}.
The phase $\psi$ is constant provided the following resonant
condition is fulfilled:
\begin{equation}
(k+\ku) \varkappa \dotl - \om = 0.
\label{dynamics.10}
\end{equation}
Using (\ref{dynamics.9}) and expressing the velocity $\dotl$ in
terms of the Lorentz factor $\gamma$ of the projectile, $\dotl = c
(1 - \gamma^{-2})^{1/2} \approx c(1 - \gamma^{-2}/2)$ (the limit
$\gamma\gg1$ is assumed), one notices that (\ref{dynamics.10})
reduces to the following relation between $\gamma$ and the wave
number $k=\om/c$:
\begin{eqnarray}
\gamma
=
\sqrt{\frac{2k}{\ku \left( 4 - a^2 k \ku \right)}}
 \equiv
\gammar.
\label{dynamics.11}
\end{eqnarray}
Resolving this equation for $k$ one
finds\footnote{In accordance with the general theory of free electron lasers (e.g.,
\cite{LuchiniMotz1990}) the amplification of the electromagnetic
wave in an undulator can occur only at the frequencies corresponding
to the harmonics of the spontaneous undulator radiation,
$\omega_k=k\, \om_1$ ($k=1,2,\dots$). The frequency $\om_1$ of the
fundamental harmonic coincides with $\om = kc$ which one finds from
(\ref{dynamics.12}). We will consider the emission stimulation
in the first harmonic only.
}
\begin{eqnarray}
k = \frac{4 \gamma_{\rm r}^2 \ku}{2 + K^2},
\label{dynamics.12}
\end{eqnarray}
where
\begin{eqnarray}
K = \gammar\,a \ku = 2\pi \gammar\,a/\lambdau
\label{dynamics.13}
\end{eqnarray}
is the undulator parameter.

It follows from (\ref{dynamics.12}) that $k \ll \ku$ in the vicinity
of the resonance (\ref{dynamics.10}). Therefore,  putting $k \pm \ku
\approx k$ in (\ref{dynamics.7}), expanding then the right-hand side
of the equation in the Fourier series and, finally, omitting the
oscillating terms one arrives at
\begin{eqnarray}
\frac{d \ps}{d t}
=
-  \frac{e E_0 a \ku}{2} \calJ(\eta) \sin \psi
\label{dynamics.15}%\label{dpldtsinJ}
\end{eqnarray}
%\begin{equation}
%\frac{d \ps}{d t} = - {e E_0 a \ku \over 2}
%\left[
%\sin \Bigl(\psi + \eta \sin (2 \ku \varkappa \ls) \Bigr)
%-
%\sin\Bigl( \psi + \varpi \sin (2 \ku \varkappa l) -2 k_{\rm u} \varkappa \ls \Bigr)
%\right]
%\label{dynamics.14}
%\end{equation}
where
\begin{eqnarray}
\eta = \frac{K^2/2}{2+K^2},
\qquad
\calJ(\eta)= J_0(\eta)-J_1(\eta),
\label{dynamics.16}%\label{varpi}
\end{eqnarray}
with $J_{0,1}(\eta)$ standing for the Bessel functions.

%The right hand side of the above equation is periodic with respect to $l$.
%The period equals to $\frac{\lambda_{\rm u}}{2 \varkappa}$.
%Assuming that the system is close to the resonance (\ref{dynamics.10})
%and therefore
%the phase $\psi$ does not vary significantly on the length
%$\frac{\lambda_{\rm u}}{2 \varkappa}$ we expand the right hand side
%into Fourier series and neglect the oscillating terms.

%{\bf Korol: to insert some connecting text.}

Let us derive the equation which describes the evolution of the
phase $\psi$ (see (\ref{dynamics.8})). Differentiating
(\ref{dynamics.3}) one gets $d \ps/d t = m \gamma^3 \ddotl \approx m
\gamma_{\rm r}^3 \ddotl$. On the other hand, double differentiation
of (\ref{dynamics.8}) yields $\ddot{\psi} = (k+\ku) \varkappa \ddotl
\approx k \ddotl$. Combining these equations with
(\ref{dynamics.15}) and noticing that $\ls\approx c t$ one derives
\begin{eqnarray}
\frac{d^2 \psi}{d \ls^2}
=
- \Om^2 \sin (\psi),
%\ddot{\psi} = - \Om^2 \sin (\psi)
%\label{dynamics.17}%\label{pendulum}
\label{dynamics.17a}%\label{penduluml}
\end{eqnarray}
which has the form of the pendulum equation.
The quantity $\Om$, defined according to
\begin{eqnarray}
\Om^2 = \frac{e E_0 k K}{2 m c^2\gamma_{\rm r}^4} \calJ(\eta),
\label{dynamics.18}%\label{Omega}
\end{eqnarray}
has the meaning of the oscillation frequency of the corresponding
simple pendulum.

The derivative $d \psi/d \ls$  can be related to the deviation of
the particle energy $\E = \gamma m c^2 $ from its  resonance value
$\E_{\rm r} = \gamma_{\rm r} m c^2 $:
\begin{equation}
\frac{d \psi}{d \ls}
\equiv \zeta
=
\frac{4 \ku}{2 + K^2} \frac{\gamma - \gammar}{\gammar}
\label{dynamics.19}%\label{zetagamma}
\end{equation}

In what follows we consider the limit of {\it small gain} and
{\it small signal}.
The limit of small gain means that the change in the amplitude $E_0$ is
much smaller than its initial value at the entrance.
In other words,  the amplitude and, consequently, the frequency
$\Om$ are approximately constant along the undulator. The small signal
limit means that $E_0$ is small enough to ensure the
inequality $\Om L \ll 1$  ($L$ is the length of the undulator).
In this case the pendulum equation (\ref{dynamics.17a}) can be solved
iteratively yielding
\begin{eqnarray}
&&
\psi(\ls) \approx  \psi_0 + \zeta_0 \ls + (\Om\ls)^2
\left(
\frac{\sin (\psi_0 +  \zeta_0 \ls)}{(\zeta_0 \ls)^2}
 -
\frac{\sin (\psi_0)}{(\zeta_0 \ls)^2}
-
\frac{\cos (\psi_0)}{\zeta_0 \ls}
\right)
\label{dynamics.20}\\%\label{psiI}\\
&&
\zeta(\ls)
\approx
\frac{\Om^2}{\zeta_0}
\Bigl(
\cos (\psi_0 +  \zeta_0 \ls)
-
\cos (\psi_0)
\Bigr)
 + \zeta_0
\label{dynamics.21}%\label{zetaI}
\end{eqnarray}
where $\psi_0$ and $\zeta_0$ denote the quantities at the undulator entrance.
%With the initial conditions
%\begin{eqnarray}
%\psi(0)  &=& \psi_0 \label{intcondpsi} \\
%\zeta(0) &=&  \zeta_0 . \label{intcondzeta}
%\end{eqnarray}
%the solution in the zero-oder approximation with respect to $\Om l$ is
%\begin{eqnarray}
%\psi \approx \psi_0 +  \zeta_0 l .
%\end{eqnarray}
%Substituting it into the pendulum equation (\ref{dynamics.17a}) and
%integration yields

%%%%%%%%%%%%%%%%%%%%%%%%%%%
\subsection{The gain factor \label{gain}}
%%%%%%%%%%%
The gain factor $\mg(L)$ characterizes the relative increase in the
energy of the amplified electromagnetic wave over the undulator length $L$.
%and during a single pass of the bunch through  the undulator.
This quantity is defined as follows
\begin{eqnarray}
 \mg(L) = \frac{\Delta\calE}{\calE(0)},
\label{gain.1}
\end{eqnarray}
where $\calE(0) = E_0^2/8\pi$ is the radiation energy density at the
entrance point, and $\Delta\calE = \calE(L) - \calE(0)$ is the
increase in the energy density (with $\calE(L)$ denoting the density
at the exit from the undulator).

It follows from the energy conservation law that the radiated energy
equals to the  decrease in the energy of the channeling particles
due to the radiative losses. Therefore, to calculate the gain factor
one can analyze the radiative energy loss of the particles.

The  energy density of the beam particles, $\calEb(\ls)$, at the
distance $\ls$  can be written as
\begin{eqnarray}
\calEb(\ls) =  \langle \E(\ls) \rangle n(\ls).
\label{gain.2}%\label{gain.2}
\end{eqnarray}
Here $n(\ls)$ stands for the volume density of the channeling particles
and $ \langle \E(\ls)\rangle$ denotes the average energy of a particle
at the distance $\ls$.
The averaging procedure takes into account that the initial beam is not
spatially modulated, i.e. that  for different particles
the instants of entry into the crystal are not correlated.
In other terms, the particles are randomly distributed
 with respect to the phase $\psi\Bigl|_{\ls=0} \equiv \psi_0$ at the
entrance point (see (\ref{dynamics.8})).
 On the other hand, (\ref{dynamics.7}) suggests that a particle's
interaction with the radiation field depends on the value of
$\psi_0$. Therefore, to obtain $ \langle \E(\ls)\rangle$ one
averages the energy $\E(\ls)=m c^2\gamma(\ls)$ with respect to
$\psi_0$: $ \langle \E(\ls)\rangle =(2 \pi)^{-1} \int_{0}^{2 \pi}
\E(\ls) d \psi_0$.
%
%The brackets mean that the quantity is averaged with respect to the
%initial phase $\psi_0$:
%\begin{eqnarray}
%\langle \dots \rangle = \frac{1}{2 \pi} \int_{0}^{2 \pi} \dots d \psi_0.
%\end{eqnarray}
%From (\ref{dynamics.8}) one sees that
%\begin{eqnarray}
%\psi_0 \equiv \psi |_{l=0} =  - \om t + \phi.
%\label{psi0def}
%\end{eqnarray}
%The initial beam is not modulated.
%Therefore, the time $t$ at which a particle enters
%the undulator is random.
%In other words, the particles in the beam are homogeneously distributed
%with respect to the initial phase $\psi_0$.
%The influence of the radiation on the particle is different
%for different $\psi_0$.
%Hence, the average with respect to $\psi_0$ has to be taken if one calculates
%the energy density (\ref{gain.2}) in the beam.

In a crystalline undulator the particles move in a medium. Due to
collisions with the crystal constituents the channeling particle
increases its transverse energy $\E_{\perp} = c^2 p_{y}^2 / 2 \E$.
At some point $\E_{\perp}$ exceeds the interplanar potential barrier
and the particle leaves the channel. The average distance traveled
by a particle in a crystal until it dechannels is called the
dechanneling length $\Ld$. In a straight crystal $\Ld$ depends on
the crystal type, the crystallographic plane, the energy and the
type of a projectile. In addition to these, $\Ld$ acquires the
dependence on the parameter $C$ \cite{KSG1999} in a bent crystal.
The dechanneling effect results in a gradual decrease in the volume
density of the channeled particles with the penetration distance.
Roughly, this decrease satisfies the exponential decay law
\cite{BiryukovChesnokovKotovBook}:
\begin{eqnarray}
n(\ls) = A\, n_0 \,  \exp(-\ls/ \Ld).
\label{gain.3}
\end{eqnarray}
Here $n_0$ is the beam density at the entrance of the crystal, and
$A$ stands for the channel acceptance (i.e., the fraction of
the incident particles which is captured into the channeling regime).

%The density $n(l)$ of {\it channeling} particles decays exponentially with $l$
%because of the dechanneling process:
%\begin{eqnarray}
%n(l) = n_0 A \exp(- l/L_{\rm d})
%\end{eqnarray}
%Here $L_{\rm d}$ is the dechanneling length,
%$n_0$ is the density of the initial beam at the entrance
%of the crystal undulator, and
%$A$ is the channel acceptance (the fraction of particle from the initial
%beam that get into the channeling regime).

It follows from (\ref{gain.2}) that the change $d \calEb$ of the
beam energy density over the interval $d \ls$ contains two terms.
One of these, proportional to the derivative of $n(\ls)$, is due to
the dechanneling process, whereas another one, proportional to $d
\langle \E(\ls) \rangle/d \ls$, describes the radiative losses. It
is exactly the latter term which is responsible for the change $d
\calE$ of the electromagnetic field energy. Therefore, one can write
\begin{eqnarray}
\frac{d \calE}{d \ls}
= - m c^2  n(\ls) \frac{d \langle \gamma \rangle}{d \ls} .
\label{gain.4}
\end{eqnarray}
%of the beam  due to
%In the expression for the derivative of ${\cal E}_{beam}(l)$ with
%respect to $l$
%\begin{eqnarray}
%\frac{d {\cal E}_{beam}}{d l} =
%m c^2  \frac{d \langle \gamma \rangle}{d l}  n(l)  +
%m c^2 \langle \gamma(l) \rangle \frac{d n}{d l} ,
%\end{eqnarray}
%only the first term is related to the interaction of the particles
% with the radiation.
%(The second one is due to the dechanneling process.)
%Therefore,
%\begin{eqnarray}
%\frac{d {\cal E}}{d l} = -
%m c^2  n(l) \frac{d \langle \gamma \rangle}{d l}
%\end{eqnarray}
%due to the energy conservation.
The derivative ${d \langle \gamma \rangle / d \ls}$  is calculated using
(\ref{dynamics.17a}) and (\ref{dynamics.19}).
Then,  integrating (\ref{gain.4}), one obtains
\begin{eqnarray}
\Delta\calE = \frac{m c^2 \gamma_{\rm r}^3}{k}\, \Om^2 \int_0^{L}
n(\ls) \langle \sin \psi(\ls) \rangle  d \ls \, .
\label{gain.5}%\label{incrE}
\end{eqnarray}
Using (\ref{dynamics.20}) and introducing the expansion in powers of
$\Om\ls \leq \Om L \ll 1$, one  carries out the averaging over $\psi_0$:
\begin{eqnarray}
\langle \sin (\psi) \rangle
=
\frac{\Om^2}{2 \zeta_0^2}
\Bigl(
 \sin (\zeta_0 \ls)  - \zeta_0 \ls \cos (\zeta_0 \ls)
\Bigr).
\label{gain.6}
\end{eqnarray}
Substituting (\ref{gain.5}) with (\ref{gain.6})  into (\ref{gain.1})
one derives the following formula for the gain factor:
\begin{eqnarray}
 \mg(L,\zeta_0)
=
8 \pi \frac{r_0 \ku}{\gamma_{\rm r}^3}\,
 \eta \calJ^2(\eta)\,
 \frac{1}{\zeta_0^2}
 \int_0^{L} n(\ls)
\Bigl(\sin (\zeta_0 \ls)  - \zeta_0 \ls \cos (\zeta_0 \ls)\Bigr) d \ls.
\label{gain.7}%\label{gL}
\end{eqnarray}
Here $r_0=e^2 / m c^2$ is the classical radius of the particle. The
second argument of the gain factor in the left-hand side  is
introduced to indicate its dependence on the quantity $\zeta_0$.

In the limit of a short crystal, $L \ll \Ld$, the dechanneling can
be neglected. Then, putting $n(\ls)\approx n(0)$ in (\ref{gain.7})
one arrives at
\begin{eqnarray}
\left.\mg(L,\zeta_0)\right|_{L \ll \Ld}
= - 2 \pi r_0 \ku \frac{L^3}{\gamma_{\rm r}^3}\, A n_0\,
 \eta \calJ^2(\eta) \,
\frac{d}{d u} \left(\frac{\sin u}{u}\right)^2
\label{gain.8}
\end{eqnarray}
where $u=\zeta_0 L / 2$.

Equation (\ref{gain.8}) presents a well-known formula for the gain
factor of the conventional free electron laser obtained within the
small signal and small gain approximation (see e.g.
\cite{LuchiniMotz1990}).

In reality, the dechanneling cannot be neglected. Indeed, the
dechanneling length of ultra-relativistic positrons (these
projectiles are the best candidates for the use in crystalline
undulators, see discussion in \cite{KSG2004_review}) measured in cm
can be roughly estimated as $\Ld \sim 0.1 \E$ with the energy in
GeV. Hence, the dechanneling length for positrons with $\E$ within
the GeV range does not  exceed several millimeters. Therefore, the
limit of a long crystal, $L\gg \Ld$ is of a great interest for an
amplifier based on a crystalline undulator. \footnote{In the
consideration we neglect the effect of the radiation attenuation in
the crystal. Therefore our results are valid if $\La \gg L$, where
$\La$ is the attenuation length.  In fact, $\La$ for the photon
energies in the range $10^2\dots 10^3$ keV is on the level of
several cm \cite{ParticleDataGroup2006}.}
%%%%%%%%%%%%%%%%%%%%%%%%%%%%%%%%%%%%%%%%%%%%%%
%\begin{figure}[tbh]
%\begin{center}
%\includegraphics[height=7cm]{gv.eps}
%\caption{The function $\frac{v}{(1+v^2)^2}$.}
% \label{figv}
%\end{center}
%\end{figure}
%%%%%%%%%%%%%%%%%%%%%%%%%%%%%%%%%%%%%%%%%%%%%%
%In fact, the dechanneling cannot be neglected in a realistic situation.
% Therefore,
%the opposite case, $L \gg L_{\rm d}$, is of greater practical importance.
%\footnote{We neglect attenuation of the
%radiation in the crystal, therefore our consideration is valid only if
%$L_a \gg L \gg L%_d$, where $L_a$
%is the attenuation length.}

Assuming in (\ref{gain.7}) $L\gg \Ld$, one extends the upper limit of
the integration to infinity and carries out the integral.
The result reads
\begin{eqnarray}
\left.\mg(L,\zeta_0)\right|_{L \gg \Ld}
=
16 \pi r_0 \ku \frac{\Ld^3}{\gamma_{\rm r}^3} A n_0 \,
 \eta \calJ^2(\eta)\,
\frac{v}{(1+v^2)^2},
\label{gain.9a}
\end{eqnarray}
where $v = \zeta_0 \Ld$.
%The function $\frac{v}{(1+v^2)^2}$ is plotted in figure \ref{figv}.
The factor ${v /(1+v^2)^2}$ attains its maximum of $3 \sqrt{3}/16$
at $v={1/\sqrt{3}}$. Therefore, in the limit $L\gg \Ld$  the maximum
gain (with respect to $\zeta_0$) is reached for
$\zeta_0=1/\sqrt{3}\Ld$. Using this  value in (\ref{gain.9a}) one
obtains the gain factor $g$ which is optimized with respect to
$\zeta_0$ and does not depend on the crystal length:
\begin{eqnarray}
g \equiv
\mg(L,\zeta_0)\Biggl|_{\genfrac{}{}{0pt}{1}{L \gg \Ld}{\zeta_0=1/\sqrt{3}\Ld}}
=
3 \sqrt{3} \pi r_0 k_{\rm u} \frac{L_{\rm d}^3}{\gamma_{\rm r}^3} A n_0
 \eta \left[ \calJ(\eta) \right]^2 .
\label{gain.9}% \label{gmaxinfty}
\end{eqnarray}
This formula coincides up to a numeric factor with the expressions obtained 
in \cite{KSG1999,KSG2004_review} within the quantum approach. The difference in 
the numeric factor is due to an additional approximation made in the cited papers.

%%%%%%%%%%%%%%%%%%%%%%%%%%%%%%%%%%%%%%%%%%%%%%%%%%%%%%%%%%%%
\subsection{Optimization of the crystal and beam parameters
\label{Optimization}}

The gain factor (\ref{gain.9}) depends on a number of parameters.
Firstly, it depends on the bending period and amplitude. Apart from
the obvious dependence of the right-hand side on the period via
$\ku=2\pi\lambdau$, the parameters $\lambdau$ and $a$ enter via
$\eta$ (see (\ref{dynamics.13}) and (\ref{dynamics.16})).
Additionally, they enter the acceptance $A$ and dechanneling length
$\Ld$ (as discussed below). The latter are also dependent on the
type of the crystal and the crystallographic plane as well as on the
particle energy $\E_{\rm r}=mc^2\gammar$.

In this section we present the scheme which allows one to define the optimum
values of these parameters, i.e. those which insure the largest
possible gain for a given crystal and initial beam density $n_0$.
Our analysis is restricted to a positron beam.

To facilitate the analysis it is convenient to express the
quantities on the right-hand side of (\ref{gain.9}) in terms of the
parameter $C$ and the undulator parameter $K$ introduced in
(\ref{dynamics.2}) and (\ref{dynamics.13}), respectively. We remind
that $C$ stands for the ratio of the maximum centrifugal force
acting on the particle in the periodically bent channel to the
maximum gradient of the interplanar potential that keeps the
particle in the channel.
%From (\ref{dynamics.1}) one sees that the maximum transverse
%acceleration of the particle is
%$\ddot{y}_{max} = a k_{\rm u}^2 \dot{z}^2 \approx a k_{\rm u}^2 c^2$.
%The maximum centrifugal force can be found by multiplying this
%expression with the effective
%mass $\gamma m \approx \gammar m$, where  $m$ is the particle rest mass.
%Therefore,
%\begin{eqnarray}
%C = \frac{\gammar m c^2 k_{\rm u}^2 a}{U'_{max}}.
%\label{C}
%\end{eqnarray}
This definition implies that  $C < 1$, otherwise all the particles will
be thrown out of the channel by the centrifugal force. The lower
limit, $C = 0$ corresponds to a straight channel, $a = 0$ (or
$\lambdau=\infty$). In this limit there is no undulator motion and,
consequently, there is no radiation amplification. Thus, the largest
gain corresponds to some optimum value of $C$ lying between $0$ and
$1$.

The $C$-dependence of the dechanneling length of an ultra-relativistic
 positron can be modeled as follows \cite{KSG1999,BiryukovChesnokovKotovBook}:
\begin{eqnarray}
\begin{cases}
\Ld  =(1-C)^{2} \Ld_0
\\
\displaystyle
\Ld_0= \gammar\, \frac{256}{9\pi^2}\,\frac{\atf}{r_0}\,\frac{d}{\Lambda},
\end{cases}
\label{Optimization.1}
\end{eqnarray}
where $\Ld_0$ stands for the dechanneling length in the straight channel,
$\atf$ is the Thomas-Fermi radius of the crystal atom, and
\begin{equation}
\Lambda =\ln(\sqrt{2 \gammar} \, mc^{2}/I)-23/24 \label{Optimization.1a}
\end{equation}
is  the 'Coulomb logarithm'
which characterizes the ionization losses of an
ultra-relativistic particle in amorphous media \cite{Landau4}
($I$ denotes an average ionization potential of the atom).

%The dechanneling length $L_{\rm d}$ depends on the factor
%$C$ \cite{BiryukovChesnokovKotovBook,KSG1999}:
%\begin{eqnarray}
%L_{\rm d} = L_{d0} (1 - C)^2.
%\end{eqnarray}
%Here $L_{d0}$ is the dechanneling length in the straight channel.

Similarly, the acceptance $A$ of a bent channel can be related to
the acceptance $A_0$ of the corresponding straight channel
\cite{BiryukovChesnokovKotovBook}:
\begin{eqnarray}
A  = (1 - C) A_0
\label{Optimization.2}
\end{eqnarray}
with $A_0 \approx 1 - 2\atf/d$.

Expressing the undulator wave number $\ku$ via $C$ and $K$
\begin{eqnarray}
\ku = \frac{C}{K}\, \frac{\dUmax}{m c^2},
\label{Optimization.3}%\label{kuCK}
\end{eqnarray}
one re-writes (\ref{gain.9}) in the form
\begin{eqnarray}
g
=
3 \sqrt{3} \pi r_0 \,
\frac{\dUmax}{m c^2}  \left(\frac{\Ld_0}{\gammar}\right)^3
n_0 \,
C (1-C)^7
\Bigl[
K^{-1}\eta \calJ^2(\eta)
\Bigr]_{\eta=K^2/(4+2K^2)}.
\label{Optimization.4}% \label{gCK}
\end{eqnarray}

The formulas (\ref{Optimization.1}) and (\ref{Optimization.1a})
suggests that the ratio $\Ld_0/\gammar$ weakly (logarithmically)
depends on the beam energy: an order of magnitude change in
$\gammar$ results in a less than 10\% change of the ratio.
Therefore, we assume this ratio to be constant for a given crystal.
Then, the gain factor, in addition to the linear proportionality
with respect to $n_0$, is a function of two independent variables,
$C$ and $K$.

The factor $C (1-C)^7$ reaches the maximum value of $7^7/8^8 \approx
0.05$ at $C=1/8$. The factor $K^{-1}\eta \calJ^2(\eta)$, as a
function of the undulator parameter $K$, attains its maximum of
$\approx 0.15$ at $K \approx 1.2$. These are the optimum values of
$C$ and $K$ which ensure the maximum gain. The latter is given by
\begin{eqnarray}
g_{\max}
=
r_0 \,
\frac{\dUmax}{m c^2}\,  \frac{\Ld^3}{\gamma_{\rm r}^3}\, n_0 .
\label{Optimization.5}%\label{gopt}
\end{eqnarray}

%%%%%% I deleted Kostyuk's figure {jmin.eps}.
% It gives nothing to the understanding of the text.
%\begin{figure}[tbh]
%\begin{center}
%\includegraphics[height=7cm]{jmin.eps}
%\caption{The function $j_{-}(K) =
%\frac{K/4}{1+K^2/2}
%[ \calJ(\eta)]^2$}
% \label{figK1}
%\end{center}
%\end{figure}
%The function $\frac{K/4}{1+K^2/2} \left[ \calJ(\eta) \right]^2$
% is shown in figure \ref{figK1}.
%It reaches its maximum value $\approx 0.15$ at $K \approx 1.2$.
%%%%%%%%%%%%%%%%%%%%%%%%%%%%%%%%%%%%%%%%%%%%%%%%%%%%%%%%%%%%%%%

The undulator period, $\lambda_{\rm u}^{\opt}$, which
corresponds to the optimal values of $C$ and $K$, one derives from
(\ref{Optimization.3}):
\begin{eqnarray}
\lambda_{\rm u}^{\opt}
\approx
60 \frac{m c^2}{\dUmax} .
\label{Optimization.6}%\label{uopt}
\end{eqnarray}

%%%%%%%%%%%% The table:
\begin{table}
\caption{Parameters of gamma ray lasers in the optimum regime for
different crystals and planes at the temperature $T=4$ K. The notation $\alpha$ stands for the
ratio $d/a \ll 1$.} \label{table1}
%\begin{tabular}{|l|l||r|r|r|r|}
\begin{tabular}{llcrrrrrrrrr}
\hline
\multicolumn{1}{c}{Crystal} & \multicolumn{1}{c}{Plane} & & \multicolumn{1}{c}{$d$} & \multicolumn{3}{c}{$\dUmax$}
& \multicolumn{1}{c}{$\lambda_{\rm u}^{\opt}$} & \multicolumn{1}{c}{$\E$}     &  \multicolumn{1}{c}{$\hbar \om$} & &
\multicolumn{1}{c}{$n_0(g_{\max}=1)$} \\
        &   &    & \multicolumn{1}{c}{\AA}   &  & \multicolumn{2}{c}{GeV/cm}
& \multicolumn{1}{c}{$\mu$m}                  & \multicolumn{1}{c}{GeV}   &
\multicolumn{1}{c}{MeV}      & &  \multicolumn{1}{c}{cm$^{-3}$} \\
\hline
C(diamond)  &  (111) & & $ 1.54 $  &    &  $  5.16 $  &  & $  59.4 $ & $  37.7 \alpha $ &  $  132 \alpha^2 $ & & $ 1.4 \cdot 10^{23} $ \\
C(graphite) &  (100) & & $ 3.35 $  &    &  $  8.77 $  &  & $  35.0 $ & $  10.2 \alpha $ &  $  165 \alpha^2 $ & & $ 5.3 \cdot 10^{21} $ \\
Si          &  (110) & & $ 1.92 $  &    &  $  4.98 $  &  & $  61.6 $ & $  31.4 \alpha $ &  $   89 \alpha^2 $ & & $ 1.2 \cdot 10^{23} $ \\
Si          &  (111) & & $ 2.35 $  &    &  $  6.28 $  &  & $  48.8 $ & $  20.4 \alpha $ &  $   47 \alpha^2 $ & & $ 4.5 \cdot 10^{22} $ \\
Ge          &  (110) & & $ 2.00 $  &    &  $ 10.94 $  &  & $  28.0 $ & $  13.7 \alpha $ &  $   37 \alpha^2 $ & & $ 7.3 \cdot 10^{22} $ \\
Ge          &  (111) & & $ 2.45 $  &    &  $ 13.55 $  &  & $  22.6 $ & $   9.1 \alpha $ &  $   20 \alpha^2 $ & & $ 2.9 \cdot 10^{22} $ \\
W           &  (110) & & $ 2.24 $  &    &  $ 40.74 $  &  & $   7.5 $ & $   3.3 \alpha $ &  $    8 \alpha^2 $ & & $ 2.0 \cdot 10^{22} $ \\
\hline
\end{tabular}
\end{table}

The optimal relativistic factor of the beam  particles $\gammar$ can
be found from (\ref{dynamics.13}) and (\ref{Optimization.3}):
\begin{equation}
\gammar = \frac{K}{a k_{\rm u}} = \frac{K^2}{C} \frac{m c^2}{a \dUmax}
\approx 11.5 \frac{m c^2}{a \dUmax}.
\label{Optimization.7}
\end{equation}
Then the energy $\hbar \omega$ of the emitted  photons can be
calculated using (\ref{dynamics.12}):
\begin{equation}
\hbar \omega = \hbar c k =
\frac{2 K^3}{C \left(1 + K^2 / 2 \right)}
\frac{\hbar m c^3}{a^2 \dUmax} =
16
\frac{\hbar m c^3}{a^2 \dUmax}.
\label{Optimization.7a}
\end{equation}
The wavelength of the produced radiation and the optimal relativistic
factor of the beam  particles $\gammar$ are not fixed by the choice
of the optimum values of the parameters $K$ and $C$. They depend on the bending
amplitude $a$. Changing $a$ while keeping the parameters $K$ and $C$
constant does not destroy the optimum regime.

The optimum values of $\lambdau$ and related parameters for
different crystals and positron beam are shown in Table
\ref{table1}. The optimum values of the beam energy $\E$ and the
photon energy $\hbar \om$ depend on the ratio $\alpha = d/a$, which
is of the order of $0.1$.

The beam density which is needed to reach $g_{\max}=1$  has been
estimated. As is seen from the last column of Table \ref{table1},
extremely high positron densities in the beam are needed to obtain a
lasing effect in a simple one-crystal amplifier even in optimized
regime. This is consistent with previous results obtained within the
quantum formalism \cite{KSG1999}.

The main reason why an appreciable gain cannot be reached at lower
densities is the fact that both the beam evolution and the emission
of output radiation take place in one crystal whose length is
limited by a few dechanneling lengths. A deeper density modulation
of the channeling beam could be obtained in a longer crystalline
undulator, but it would not lead to the increase of the radiation
gain. The reason is the exponential decay of the density of the
radiating particles because of the dechanneling process. This
dilemma is resolved in the next section where the scheme of a
two-crystal gamma ray amplifier - the gamma klystron - is presented.

%%%%%%%%%%%%%%%%%%%%%%%%%%%
\section{Two-crystal gamma ray amplifier --- the klystron}\label{two-crystal-amp}

\begin{figure}[tb]
\includegraphics[width=\textwidth]{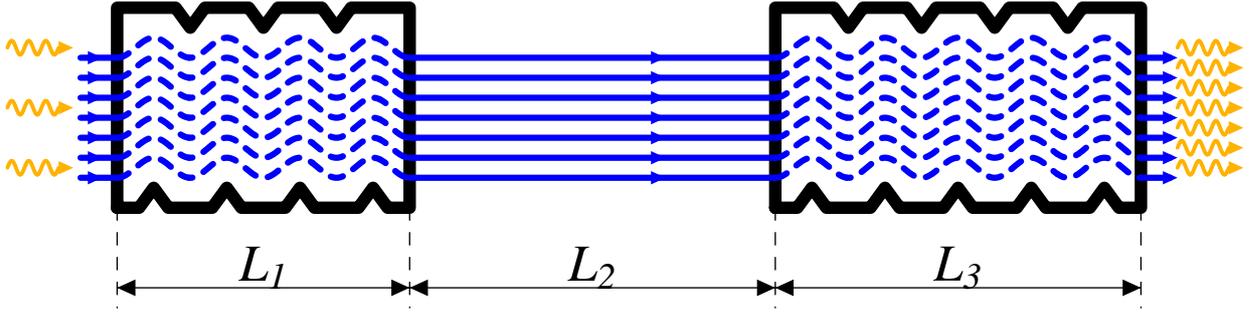}
\caption{A scheme of the gamma klystron. A beam of charged particles
(solid lines) and initial (seed) radiation (the solid wavy lines)
enter the first crystal with periodically bent crystallographic
planes of the length $L_1$. The particle follow the shape of the
crystallographic planes and move along nearly sinusoidal
trajectories (the wavy dashed lines) inside the crystal. Interaction
between the seed radiation and the beam in the first crystal gives
rise to a correlation between the particle momentum and its position
along $z$ axis. This correlation is transformed into a modulation of
the beam density  while the beam travels in the vacuum gap of the
length $L_2$. Then output radiation is produced by density-modulated
beam in the second crystal of the length $L_3$ (see text for
details). } \label{klystron}
\end{figure}

The scheme of the gamma klystron is shown in figure \ref{klystron}.
Two periodically bent crystals of the lengths $L_1$ and $L_3$,
respectively, are separated by a vacuum gap of the length $L_2$. The
beam of particles  passes the both crystals successively. A
correlation between the particle momentum and its position along the
beam direction is created due to the interaction of the channeling
particle with the seed radiation in the first crystal. This
correlation is further transformed into the density modulation of
the beam in the vacuum gap. Production of the output radiation takes
place in the second crystal.

The idea of the crystal undulator based gamma klystron is very
similar to that of optical klystron
\cite{Vinokurov,VinokurovSkrinsky}.

%%%%%%%%%%%%%%%%%%%%%%%%%%
\subsection{Particle dynamics in the Gamma Klystron}

In the first undulator, the phase evolves according to
(\ref{dynamics.17a}) and the solution is given  by
(\ref{dynamics.20}) and (\ref{dynamics.21}) so that the following
expressions are valid at exit from the first undulator
\begin{equation}
\left \{
\begin{array}{rcl}
\zeta_1 & \approx & \zeta_0 + \frac{\Om^2}{\zeta_0} \cos (\psi_0 +
\zeta_0 L_1) -
\frac{\Om^2}{\zeta_0} \cos (\psi_0) \, ,  \\
\psi_1 &\approx&  \psi_0 + \zeta_0 L_1 + ( \Om L_1 )^2 \left [
\frac{\ds \sin (\psi_0 +  \zeta_0 L_1)}{\ds (\zeta_0 L_1)^2}
 - \frac{\ds \sin (\psi_0)}{\ds (\zeta_0 L_1)^2} - \frac{\ds \cos (\psi_0)}{\ds \zeta_0 L_1}
\right ] \, .
\end{array}
\right .
\label{zetapsi1}
\end{equation}

In vacuum, the particles move along straight trajectories.
Therefore, the the equation of motion reduces to $ d^2 \psi / d
\ls^2 = 0 $. Hence,
\begin{equation}
\left \{
\begin{array}{rcl}
\zeta_2 & = & \zeta_1 \, ,  \\
\psi_2 & = & \psi_1 + \zeta_1 L_2
\end{array}
\right . \label{zetapsi2}
\end{equation}
at entrance of the second undulator.

In the second undulator, the phase evolves again according to (\ref{dynamics.17a}) leading to
\begin{equation}
\left \{
\begin{array}{rcl}
\zeta &\approx& \zeta_2 + \frac{\ds \Om^2}{\ds \zeta_2} \cos (\psi_2 +
\zeta_2 s) -
\frac{\ds \Om^2}{\ds \zeta_2} \cos (\psi_2)  \, ,   \\
\psi &\approx&  \psi_2 + \zeta_2 s + ( \Om s)^2 \left [ \frac{\ds  \sin
(\psi_2 +  \zeta_2 s)}{\ds (\zeta_2 s)^2}
 - \frac{\ds \sin (\psi_2)}{\ds (\zeta_2 s)^2} - \frac{\ds \cos (\psi_2)}{\ds \zeta_2 s}
\right ] \, ,
\end{array}
\right .
\label{zetapsi3}
\end{equation}
where $s$  is measured from the entrance of the second undulator, $0 \leq s\leq L_3$.

We assume that
\begin{equation}
L_2 \gg L_1,L_3.
\label{L2ggL1L3}
\end{equation}
Therefore,  $\Om L_2$ or $\Om^2 L_1 L_2$ have not to be small even
though $\Om L_1 \ll 1$ and $\Om s \leq \Om L_3 \ll 1$.
Hence, neglecting the terms proportional to $\Om L_1$ and
$\Om s$, but keeping the $\Om L_2$ terms one derives
\begin{eqnarray}
\sin \psi = \sin \left \{ \psi_0 + \zeta_0 (L_1 + L_2 + s) + \Om^2
L_1 L_2 \frac{\cos (\psi_0 +  \zeta_0 L_1) - \cos (\psi_0)}{\zeta_0
L_1} \right \} .
\end{eqnarray}
Averaging with respect to $\psi_0$ leads to:
\begin{eqnarray}
\langle \sin \psi \rangle  &=&  - J_1 \left [ 2
\frac{\Om^2 L_2}{\zeta_0} \sin \left( \frac{\zeta_0 L_1}{2} \right)
\right ] \sin \left[ \zeta_0 \left( \frac{L_1}{2} + L_2 + s \right)
\right] . \nonumber
\end{eqnarray}

\subsection{Radiation Gain in the Gamma Klystron}

The increment of the radiation energy density in the
second\footnote{ Utilizing of the klystron scheme makes sense only
if the radiation gain in second crystal is much larger than in the
first one. Therefore the gain in the first crystal is neglected.
% The radiation gain in the first crystal is the same as in the
% one-crystal amplifier (\ref{gain.7}).
} undulator is found as (cf.
(\ref{gain.5}))
\begin{eqnarray}
\Delta{\cal E} =
 \frac{m c^2 \gamma_{\rm r}^3}{k} \Om^2 \int_0^{L_3} n(s) \langle \sin \psi(s) \rangle  d s 
 \, ,
\label{incrE3}
\end{eqnarray}
where $n(s)$ stands for the volume density of channeling particles
in the second undulator. To calculate $n(s)$, one has to take into
account the channel acceptance $A$ twice: at the entrances of the
first and the second undulator\footnote{Strictly speaking the
acceptances of two undulators are not exactly equal because of
different angular distributions of particles at the entrances of the
undulators. But this difference is small and is neglected in our
calculations.}. Additionally, the decrease in the number 
of channeling particles  due to
dechanneling in both undulators has to be taken into consideration.
Therefore, the density of channeling particles in the second
undulator reads
\begin{eqnarray}
n(s) = A^2 \exp \left( - \frac{L_1 + s}{L_{\rm d}}  \right) n_0 .
\end{eqnarray}
Here $n_0$ is density of the particles at the entrance of the first
undulator.

We consider the case $L_3 \gg L_{\rm d}$ so that the upper limit of the integral in (\ref{incrE3})
can be replaced by the infinity. One obtains after the integration
\begin{equation}
\Delta \mathcal{E} = - \frac{m c^2 \gamma_{\rm r}^3}{k} \Om^2 A^2
n_0 \exp \left( - \frac{L_1}{L_{\rm d}}  \right) J_1 \left (
\Om^2 L_1 L_2 \frac{\sin u_1}{u_1} \right )
% \nonumber \\
% & &
L_{\rm d} \frac{\sin \left[ u_1 + 2 u_2 + \arctan (v) \right]}
{\sqrt{1 + v^2}} ,
\label{incrE3i}
\end{equation}
where $u_i = \zeta_0 L_i / 2$, $i=1,2$ and $v = \zeta_0 L_{\rm d}$.

The above expression is derived in the small-gain approximation,
but not in the small-signal approximation, i.e.  it is valid even
when the argument of the Bessel function is of the order of one. In
the following, however, we restrict our consideration to the weak
signal regime to make a comparison with the one-crystal
amplifier.

In the small-signal case, the radiation gain does not depend on the
amplitude of the initial wave and has the form
\begin{equation}
\mathfrak{g}(L_1, L_2, \infty) = \frac{\Delta \mathcal{E}}{\mathcal{E}}  =
 8 \pi r_0 k_{\rm u} \frac{L_1 L_2 L_{\rm d}}{\gamma_{\rm r}^3} A^2 n_0
\exp \left( - \frac{L_1}{L_{\rm d}}  \right)
\eta \left[ \calJ(\eta) \right]^2
h (u_1, u_2, v)
\end{equation}
with
\begin{equation}
h (u_1, u_2, v) = - \frac{\sin u_1}{u_1} \frac{\sin \left[ u_1 + 2
u_2 + \arctan (v) \right]} {\sqrt{1 + v^2}} \, .
\end{equation}

The optimum value of $L_1$ can be chosen by maximizing the function
$L_1 \exp \left( - L_1/L_{\rm d}  \right)$. The maximum is reached at $L_1
= L_{\rm d}$.

For fixed values of $L_1$, $L_2$ and $L_3$, the variables $u_1$,
$u_2$ and $v$ depend only on single variable, $\zeta_0$. The
optimum value of $\zeta_0$ is found by maximizing the function $- h
(u_1, u_2, v)$.  Taking into account that $u_2 \gg u_1, v$ due to
(\ref{L2ggL1L3}) one finds that the maximum is reached at $u_2
\approx - \pi/4$ and is approximately equal to $1$.

% For illustrative purposes, we have shown the behavior of the function \linebreak
% $- \frac{\sin u_1}{u_1} \frac{\sin \left[ u_1 + 2 u_2 + \arctan (v) \right]}
% {\sqrt{1 +v^2}} $
% for the case $L_2 = 10 L_{\rm d} = 10 L_1$. In this case $u_1 = u_2/10$ and $v=u_2/5$.
% \begin{figure}[bht]
% \begin{center}
% \includegraphics[height=7cm]{hu2.eps}
% \caption{The function $h(u2) = - \frac{\sin u_1}{u_1}
% \frac{\sin \left[ u_1 + 2 u_2 + \arctan (v) \right]}{\sqrt{1 + v^2}} $
% at $u_1 = u_2/10$ and $v=u_2/5$.}
%  \label{fig3}
% \end{center}
% \end{figure}
% As is seen, the maximal value of this function is approximately $1$ and is reached at

Therefore the maximum radiation gain which can be reached in the
gamma klystron is
\begin{equation}
g \equiv  \left. \mathfrak{g}(L_{\rm d}, L_2, L_3) \right |_{\genfrac{}{}{0pt}{1}{ L_3 \ll L_{\rm d}}
{\zeta_0 = - \pi / (2 L_2) }} = \frac{ 8 \pi}{\rm e} r_0 k_{\rm u}
\frac{L_{\rm d}^2 L_2}{\gamma_{\rm r}^3} A^2 n_0  \eta \left[ \calJ(\eta)
\right]^2 .  \label{gklystron}
\end{equation}
Here ${\rm e} = 2.7182818\dots$ is the base of the natural
logarithm.\footnote{The constant ${\rm e}$ in (\ref{gklystron}) and
(\ref{ratio}) should not be confused with the particle charge $e$.}

Comparing
this formula with (\ref{gain.9}) one sees that, for the same
parameters of the crystals and the beam, the radiation gain in the
gamma klystron exceeds the gain achievable in  the one-crystal amplifier
by the following factor
\begin{eqnarray}
\frac{g (\mbox{klystron})}{g (\mbox{one-crystal})} = \frac{8}{3
\sqrt{3} {\rm e}}  A \frac{L_2}{L_{\rm d}} .
\label{ratio}
\end{eqnarray}
This means that a significant gain can be obtained in the gamma klystron at much lower beam densities
than in the one-crystal amplifier, provided that $L_2$ is essentially larger
than $L_{\rm d}$.

\section{Multicascade amplifier}\label{multicascade}

So far we considered a hard X ray or gamma ray amplifier in the small gain regime. Estimations in 
section \ref{Optimization} were made for the case $g=1$, i.e. when the amplifier doubles the intensity
of the electromagnetic radiation. Of cause, such a small gain is insufficient. A usefull device should 
be able to amplify the signal by several orders of magnitude. There are two methods of increasing 
the gain. The first one is straightforward. It is merely increasing the density of the positron beam. 
This will switch the amplifier into the large gain regime. The potential of this method is, however, 
limited. For one crystal amplifier, one needs very high positron densities, $n_0 \sim 10^22$ cm$^{-3}$, 
to reach even $g=1$ . Using the klystron scheme will decrease the minimum necessary density  by severals orders
of magnitude. Nonetheless, this quantity will likely remain at the edge of the capabilities of the accelerator
technique and the sustainability of the crystalline materials. Therefore, increasing the positron density much 
beyond the minimum necessary level will be difficult. 

\begin{figure}[thb]
\includegraphics[width=\textwidth]{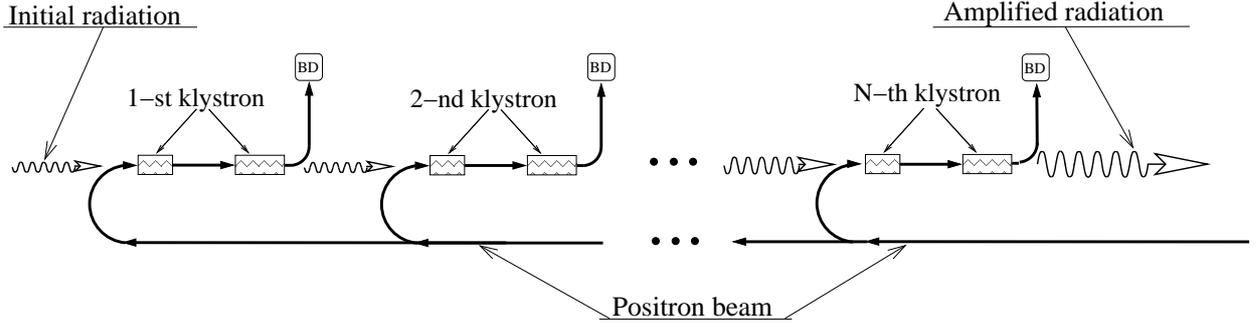}
\caption{A scheme of a multicascade hard X ray or gamma ray amplifier. 
Each cascade consist of a klystron fed with a separate positron 
bunch. The radiation passes the klystron successively.
If the gain factor 
of each cascade is $g$, the radiation intensity will be amplified 
by the factor of $G=(1+g)^{N}$, where $N$ is the number of the 
cascades. "BD" stands for the positron beam dump.}
\label{multicascade_fig}
\end{figure}

This difficulty can be overcome by using the 
second method of increasing the gain: combining several klystrons into a multicascade amplifier. This 
scheme is shown in figure \ref{multicascade_fig}. Scattering the beam particles in the crystal decreases the
beam intensity and increases its emittance. Thereofore, each cascade should be fed by a 'fresh' positron bunch from 
the accelerator. The radiation passes each cascade successively and is being gradually amplified. 
The distance between the bunches has to be chosen in such a way that each bunch 
enters the crystal at the same time as the wavepackage of the radiation reaches the corresponding cascade.
The resulting 
amplification factor will be $G=(1+g)^{N}$, where $g$ is the gain of each cascade and $N$ is the number of cascades.

\section{Conclusion}\label{conclussion}

We analyzed two different schemes of the gamma ray amplifier. Both
schemes are based on crystalline undulators - single crystals with
periodically bent crystallographic planes exposed to a high energy charged
particle beam. Due to the channeling phenomenon,
 ultra-relativistic charged particles move along
nearly sinusoidal trajectories inside the crystal. Initial gamma  radiation
traveling through the crystal parallel to the beam is amplified due to
interaction with the channeling particles.

We have shown, that the simplest one-crystal scheme would require
extremely high (more than $10^{21}$ particles per cubic centimeter)
positron beam densities to obtain significant amplification. Which
makes practical realization of this scheme extremely challenging.

On the other hand, the two-crystal klystron scheme seems to be more
promising. A significant gain could be obtained at much lower densities, provided that
the distance $L_2$ between two crystals in the klystron
exceeds greatly the dechanneling length $L_{\rm d}$. The question how large 
the distance $L_2$  can be in a realistic
device cannot be answered within the present approach. 
The simple one-dimensional 
model considered in the present paper does not put any restrictions on the 
distance $L_2$. In reality, the
restrictions on $L_2$ come from the energy spread (longitudinal
temperature) of the beam. This is, however a technical restriction that depends 
on the quality of the beam determined by the parameters of the accelerator. 
Physical restrictions are more important. They appear, for instance, due to the 
longitudinal velocity spread 
induced  by the channeling oscillations and incoherent 
scattering in the first crystal or the  beam divergence in the
vacuum gap due to the volume charge.

Hence further analysis is needed to give the final answer
about the feasibility of practical realization of the gamma klystron.
%It will be presented in our future publications.

%%%%%%%%%%%%%%%%%%%%%%%%%
\section{Acknowledgments}

This work has been supported by the European Commission
(the PECU project, Contract No. 4916 (NEST)).

\end{document}